\documentclass[1p]{elsarticle}
\biboptions{sort&compress}
\journal{Nuclear Physics A}
\usepackage{graphicx}
\usepackage{amsfonts}
\usepackage{graphicx}
\usepackage{amsmath}
\usepackage{mathtools}
\usepackage{amssymb}
\usepackage{pmat}
\usepackage{ifthen}
\usepackage{isotope}

\usepackage{liealg}
\usepackage{wigner}

\newcommand{\Nex}{{N_\text{ex}}}  % double braces to help with subscripting on W, but impede superscript
\newcommand{\NexBARE}{N_\text{ex}} %...to allow superscript
\newcommand{\Nexhat}{{\hat{N}_\text{ex}}} 
\newcommand{\Nrel}{N_\text{rel}}
\newcommand{\Nmax}{N_\text{max}}
\newcommand{\Ncm}{N_\text{c.m.}}
\newcommand{\Hcm}{H_\text{c.m.}}
\newcommand{\grpsutimes}{\grpsu{3}\times\grpsu{2}}

\newcommand{\Bdcmlm}[1][]{B^{\dagger(10)}_{\mathrm{c.m.}\ifthenelse{\equal{#1}{}}{}{,}#1}}
\newcommand{\Btcmlm}[1][]{\tilde{B}^{(01)}_{\mathrm{c.m.}\ifthenelse{\equal{#1}{}}{}{,}#1}}

\newcommand{\Bdcm}[1][]{B^{\dagger}_{\mathrm{c.m.}\ifthenelse{\equal{#1}{}}{}{,}#1}}
\newcommand{\Btcm}[1][]{\tilde{B}_{\mathrm{c.m.}\ifthenelse{\equal{#1}{}}{}{,}#1}}

\newcommand{\inline}[1]{\smash{\cramped{#1}}}

\newcommand{\proton}{p}
\newcommand{\neutron}{n}
\newcommand{\Nsp}{\eta}
\newcommand{\sNsp}{{s,\Nsp}}

\usepackage{stackrel}
\DeclareRobustCommand{\subqnstrut}{\makebox[0pt]{\phantom{()}}}
\DeclareRobustCommand{\subqn}[2]{\stackrel[{#2\subqnstrut}]{}{#1\subqnstrut}}

\DeclareRobustCommand{\falling}[2]{{{#1}^{\underline{#2}}}}

\newcommand{\wfgen}{\psi} % single bracket permits superscript
\newcommand{\wfgencm}[1][]{\psi_{\text{c.m.}#1}}
\newcommand{\wfgenin}[1][]{\psi_{\text{in}#1}}

\begin{document}
%bibliographystyle{iopart-num}
%bibliographystyle{apsrevm}

\title{Construction of the center-of-mass free space\\ for the $\grpsu{3}$ 
      no-core shell model}

%% \author{F Q  Luo$^1$, M A Caprio$^1$ and T Dytrych$^2$}
%% \address{${^1}$Department of Physics, University of Notre Dame, Notre Dame, IN 46556-5670, USA}
%% \address{${^2}$Department of Physics and Astronomy, Louisiana State University, Baton Rouge, LA 70803-4001, USA}

\author[nd]{F. Q. Luo}
\author[nd]{M. A. Caprio}
\author[lsu]{T. Dytrych}
\address[nd]{Department of Physics, University of Notre Dame, Notre Dame, IN 46556-5670, USA}
\address[lsu]{Department of Physics and Astronomy, Louisiana State
University, Baton Rouge, LA 70803-4001, USA}

\date{\today}

\begin{abstract}
We address the removal of states with center-of-mass excitation from
the $\grpsu{3}$ no-core shell model [$\grpsu{3}$-NCSM] space,
\textit{i.e.}, construction of the nonspurious subspace.  A procedure
is formulated based on solution of the null-space problem for the
center-of-mass harmonic oscillator lowering operator
$\inline{\Btcmlm}$, operating at the level of $\grpsu{3}$ irreducible
representations.  Isolation of the center-of-mass free subspace for
the $\grpsu{3}$-NCSM provides the foundation for exact removal of
center-of-mass dynamics in the proposed $\grpsptr$ symplectic
no-core shell model.  We outline the construction
process for the matrix representation of $\inline{\Btcmlm}$, present
the algorithm for obtaining the nonspurious space, and examine the
dimensions obtained for center-of-mass free $\grpsu{3}$ subspaces in
representative light nuclei.
\end{abstract}

\maketitle

%%PACS: 21.60.Cs, 21.60.Fw
%%\pacs{21.60.Cs,21.60.Fw}
% 21.10.-k properties of nuclei; nuclear energy levels
% 21.10.Ky electromagnetic moments
% 21.10.Ft charge distribution
% 21.10.Re collective levels
% 21.30.-x nuclear forces
% 21.45.-v few-body systems
% 21.60.De collective models
% 21.60.Cs shell model 
% 21.60.Fw models based on group theory
% 27.20.+n 6<=A<=19
% 02.30.Gp special functions

\section{Introduction}
\label{sec-intro}

The ability to carry out no-core configuration interaction
calculations of light nuclei, in the no-core shell model
(NCSM)~\cite{navratil2000:12c-ab-initio,navratil2000:12c-ncsm,navratil2009:ncsm},
has made a significant contribution to recent progress in the
\textit{ab initio} description of nuclei.
However, the dimensionality of the nuclear model space becomes
computationally prohibitive as the number of active nucleons and
orbitals increases.  Symmetry can play a significant role in
addressing this problem, by assisting in the selection of the
physically relevant portions of the model space.
Elliott~\cite{elliott1958:su3-part1,elliott1958:su3-part2,elliott1963:su3-part3}
explored the $\grpu{3}\supset\grpsu{3}$ symmetry of the harmonic
oscillator, which serves as an organizational scheme for quadrupole
deformation and rotation in the nuclear shell model.  The $\grpu{3}$
algebra of the oscillator is contained in a larger $\grpsptr$ algebra,
which is found to have a close connection both to the dominant
components of the nuclear Hamiltonian and to nuclear collective
motion~\cite{rosensteel1977:sp6r-shell,rosensteel1980:sp6r-shell,draayer1984:spsm-20ne}.
Following the discovery of evidence for $\grpsu{3}$ and $\grpsptr$
symmetry in the nuclear eigenstates obtained in conventional NCSM
calculations for light
nuclei~\cite{dytrych2007:sp-ncsm-evidence,dytrych2007:sp-ncsm-dominance},
the NCSM has been reformulated in terms of an $\grpsu{3}$-based model
space, in the $\grpsu{3}$ no-core shell model
[$\grpsu{3}$-NCSM]~\cite{dytrych2012:su3ncsm-hites12,dytrych:inprep}.
This development provides the foundations for future realization of a
symplectic no-core shell model (Sp-NCSM)~\cite{dytrych2008:sp-ncsm},
making full use of $\grpsptr$ symmetry.

In predicting physical properties of the nucleus, only the intrinsic
dynamics of the nucleus is of interest, not the center-of-mass motion.
In principle, the center-of-mass motion may be eliminated from the
problem by explicitly changing variables to relative coordinates.
However, the nuclear many-body state must be antisymmetrized, and with
increasing nucleon number this process rapidly becomes prohibitive in
relative coordinates~\cite{navratil2009:ncsm}.  On the other hand, for
a many-body basis constructed from antisymmetrized products of
single-particle states, as in the NCSM, antisymmetrization is
straightforward, but both center-of-mass (spurious) and intrinsic
excitations are included in the model
space~\cite{elliott1955:com-shell,baranger1961:shell-com}.
Consequently many of the nuclear eigenstates obtained by diagonalizing
the Hamiltonian will carry center-of-mass excitation.  Moreover, exact
factorization of the many-body wave function into center-of-mass and
intrinsic parts~--- \textit{i.e.},
$\wfgen(\vec{r}_i;\vec{\sigma}_i)=\wfgencm(\vec{R})\wfgenin(\vec{r}_{ij};\vec{\sigma}_i)$,
where $\wfgencm(\vec{R})$ depends on the center-of-mass coordinate,
and $\wfgenin(\vec{r}_{ij};\vec{\sigma}_i)$ depends on relative
coordinates $\vec{r}_{ij}$ and intrinsic spin degrees of freedom~---
is obtained only within certain specific truncations of the model
space.  In the context of the conventional NCSM, for which the basis
is constructed from antisymmetrized products of oscillator wave
functions, factorization is obtained if truncation is carried out
according to the $\Nmax$ scheme, that is, by the total number of
oscillator quanta for the many-body state~\cite{navratil2009:ncsm}.
Within the $\grpsu{3}$-NCSM, factorization is also retained in model
spaces which have furthermore been truncated according to $\grpsu{3}$
and spin quantum numbers (see Section~\ref{sec-cmf}).  Otherwise,
factorization of intrinsic and center-of-mass wave functions is in
general incomplete, in which case all eigenstates contain some
spurious contribution, presenting challenges for the study of the
intrinsic excitations (\textit{e.g.},
Refs.~\cite{lipkin1958:com-shell,mcgrory1975:spurious-com,caprio2012:csbasis}).

The usual approach to addressing the problem of spurious states
consists of modifying the nuclear \textit{Hamiltonian}, working within a model
space which supports factorization, by adding a
Lawson term~\cite{gloeckner1974:spurious-com} proportional the
harmonic oscillator Hamiltonian $\Hcm$ for the center-of-mass
coordinate.  After diagonalization, the spurious states remain in the
spectrum but are shifted to an excitation energy above the low-lying
intrinsic states of interest.  However, the other possibility is to
modify the \textit{model space}, so as to explicitly remove the entire space of
spurious states, before the Hamiltonian is diagonalized.  This
approach may be made feasible through the use of $\grpsu{3}$
symmetry~\cite{kretzschmar1960:su3-shell-part2-com,verhaar1960:shell-com,hecht1971:su3-com,millener1975:14b-beta-su3,millener1992:su3-multi-shell}.
After the spurious states are removed, the remaining center-of-mass
free (CMF) states form a suitable model space for
describing the intrinsic dynamics of the nucleus.

The CMF states are associated with the lowest eigenvalue of the
center-of-mass Hamiltonian $\Hcm$ and thus possess a harmonic
oscillator $0s$ wave function in the center-of-mass degrees of
freedom.  They also therefore constitute the null space of the
center-of-mass harmonic oscillator lowering operator
$\inline{\Btcmlm}$.  The approach developed here for isolating the CMF
space is based on solution of the null-space problem for this operator,
formulated at the level of subspaces of definite $\grpsu{3}$ symmetry and intrinsic spin.  

A principal motivation is to enable exact separation of center-of-mass
dynamics in extensions to the $\grpsu{3}$-NCSM, in particular the
Sp-NCSM.  The many-body basis for the $\grpsu{3}$-NCSM is
the starting point for defining the basis for the
Sp-NCSM~\cite{dytrych2008:sp-ncsm}.  Briefly, an $\grpsptr$
irreducible representation (irrep) is constructed from an extremal
$\grpsu{3}$ state~\cite{dytrych2008:sp-ncsm}, \textit{i.e.}, possessing the fewest
oscillator quanta. Other states within the irrep are constructed by
acting on this extremal state with an $\grpsu{3}$-coupled product
$[A^{(20)}\times
A^{(20)}...\times A^{(20)}]^{(\lambda\mu)}$, where
$A^{(20)}$ is the translationally-invariant $\grpsptr$ raising
operator.  Each application of $A^{(20)}$ contributes two
oscillator quanta to the system.  Since the action of $A^{(20)}$
does not introduce center-of-mass excitation, an $\grpsptr$ irrep is
free of center-of-mass excitation as long as it is built from an
extremal $\grpsu{3}$ state which is free of such excitation.  The
present results therefore provide the foundation for obtaining the CMF
model space for the Sp-NCSM.

The construction of the $\grpsu{3}$-NCSM basis, before removal of
spurious contributions, is first outlined
(Section~\ref{sec-su3}).  We then introduce an algorithm for
identification of the CMF subspace within the $\grpsu{3}$-NCSM model space,
based on constructing the
matrix representation of $\inline{\Btcmlm}$ between
$\grpsu{3}$-coupled subspaces and solving for its null space
(Section~\ref{sec-cmf}).  Finally, we summarize the dimensions
obtained for CMF $\grpsu{3}$ subspaces in
representative light nuclei (Section~\ref{sec-calc}).  Preliminary
results were reported in Ref.~\cite{luo2012:su3cmf-hites12}.

\section{\boldmath $\grpsu{3}$-NCSM basis states}
\label{sec-su3}

The many-body basis states for the $\grpsu{3}$-NCSM have good
$\grpsutimes$ quantum numbers, where the $\grpsu{3}$ symmetry label
$(\lambda\mu)$ characterizes the spatial degrees of freedom, according
to the Elliott classification, while the $\grpsu{2}$ label $S$ describes
the total intrinsic spin angular momentum.  The creation operator for
a nucleon in a given major oscillator shell $\Nsp$ comprises an
$\grpsutimes$ tensor $\inline{a^{\dagger}_{(\Nsp 0)1/2}}$, where the
labels denote
$(\lambda\mu)S=(\Nsp0)\tfrac12$~\cite{escher1998:sp6r-shell-su3coupling}.
The operators $\inline{a^{\dagger}_{(\Nsp 0)1/2}}$ are then used as
the fundamental units in building up an $\grpsutimes$-coupled nuclear
state.

Specifically, each $\grpsu{3}$-NCSM basis state is characterized by a
definite distribution of nucleons over the major shells.  First, all
nucleons in the each major shell $\Nsp$ are combined to form a
configuration of the type $\inline{[a^{\dagger}_{(\Nsp 0)1/2}\times
a^{\dagger}_{(\Nsp 0)1/2} \times
\cdots]^{(\lambda_{\Nsp}\mu_{\Nsp})S_{\Nsp}}}$, with
good $\grpsutimes$ coupling~\cite{draayer1989:un-u3-plethysm},
separately for the protons and neutrons.  Then, such configurations
for individual major shells are coupled successively to form a total
proton state and total neutron state carrying good $\grpsutimes$
quantum numbers. Finally, the proton state and neutron state are
coupled, to give the $\grpsu{3}$-NCSM basis state.  Since the major
shells have definite occupations, the state may be classified, as
usual in the oscillator basis for the NCSM~\cite{navratil2009:ncsm},
by the number $\Nex$ of harmonic oscillator excitations, taken
relative to the minimal number of oscillator quanta possible for the
given number of protons and neutrons.

The resulting $\grpsu{3}$-NCSM basis state has the form, with all
coupling labels shown explicitly,
\begin{multline}\label{eqn:SU3xSU2_basis_states}
\Big( \tket{ ( ( ( ( \gamma_{\proton,0} \times\gamma_{\proton,1}
)^{\rho_{\proton,0}\omega_{\proton,0}} \times \gamma_{\proton,2} )^{
\rho_{\proton,1}\omega_{\proton,1}} \times \gamma_{\proton,3} )^{
\rho_{\proton,2}\omega_{\proton,2}} ... \times \gamma_{\proton,\Nsp_\text{max}}
)^{\rho_{\proton}\omega_{\proton} }}
\\
\times  
 \tket{ ( ( ( ( \gamma_{\neutron,0} \times\gamma_{\neutron,1}
)^{\rho_{\neutron,0}\omega_{\neutron,0}} \times \gamma_{\neutron,2} )^{
\rho_{\neutron,1}\omega_{\neutron,1}} \times \gamma_{\neutron,3} )^{
\rho_{\neutron,2}\omega_{\neutron,2}} ... \times \gamma_{\neutron,\Nsp_\text{max}}
)^{\rho_{\neutron}\omega_{\neutron}} }
\Big)^{\rho\omega}.  
\end{multline}
Here the symbol $\gamma_\sNsp$ represents the
labels $\gamma=[f_1,f_2,\ldots,f_N]\alpha(\lambda\mu)S$ needed to
completely specify the coupling
of nucleons of type $s$ (\textit{i.e.}, protons or neutrons) within
major shell $\Nsp$~\cite{draayer1989:un-u3-plethysm}.  Specifically, each major
shell has associated with it a $\grpu{N}$ algebra
[$N=(\Nsp+1)(\Nsp+2)$] consisting of bilinears of creation and
annihilation operators, for which the irreps are
labeled by $[f_1,f_2,\ldots,f_N]$, where we consider only antisymmetric irreps, and 
$\sum_{i=1}^N f_i$ equals the occupation of the shell.  Within a $\grpu{N}$ irrep, a multiplicity index $\alpha$
is required to distinguish $\grpsutimes$ irreps with the same quantum numbers
$(\lambda\mu)S$, yielding the labeling scheme
\begin{equation}
\subqn{\grpu{N}}{[f_1,f_2,\ldots,f_N]}  \subqn{~~\supset~~}{\alpha}   \subqn{\grpsu{3}}{(\lambda\mu)}\times\subqn{\grpsu{2}}{S}.
\end{equation}
The symbols $\omega_\sNsp$ in~(\ref{eqn:SU3xSU2_basis_states}) then  
indicate the $\grpsutimes$ coupling labels [$\omega\equiv(\lambda\mu)S$] of successive
shells, and $\rho_\sNsp$ denotes the multiplicity index for this coupling.
Finally we have total couplings
$\omega_{\proton}\equiv (\lambda_{\proton}\mu_{\proton})S_{\proton}$ for the protons,
$\omega_{\neutron}\equiv (\lambda_{\neutron}\mu_{\proton})S_{\neutron}$ for the neutrons, and
$\omega\equiv (\lambda\mu)S$ for the entire basis state, with corresponding multiplicity indices $\rho_\proton$, $\rho_\neutron$, and $\rho$, respectively.

The expression in~(\ref{eqn:SU3xSU2_basis_states}) represents not
just a single state but an entire set of states, with various values for the quantum numbers 
associated with the branching of $\grpsutimes$ into angular momentum
subalgebras: the orbital (spatial) angular momentum $L$, the inner multiplicity
$\kappa$ for this $L$ within the $\grpsu{3}$ irrep $(\lambda\mu)$, and
the total angular momentum $J$, as well as its $z$-projection $M$.
However, these states share the same ``internal'' microscopic structure,
given by the same couplings of the particles at the level of
$\grpsutimes$.  Therefore, they may be thought of as a single
\textit{reduced state} for certain purposes, in particular, evaluation of
reduced matrix elements under the $\grpsutimes$ Wigner-Eckart theorem and, as we shall see,
identification of CMF linear combinations.
The analogy is to
angular momentum theory, where one may consider the states
$\tket{JM}$, for different $M$, to be substates of a single state
$\trket{J}$, more formally, a tensorial set or $\grpsu{2}$ irrep.  

\section{Construction of the CMF subspace}
\label{sec-cmf}

The separation of the many-body space into CMF and spurious parts
simplifies in the context of an $\grpsutimes$-coupled basis, since the
process may be carried out independently within subspaces
characterized by definite $\grpsutimes$ quantum numbers.  To start
with, the center-of-mass Hamiltonian $\Hcm$ does not connect states
involving different numbers $\Nex$ of oscillator excitations, \textit{i.e.}, $[\Hcm,\Nexhat]=0$.  It is
due to this property that the usual $\Nmax$ truncation scheme for the
NCSM~\cite{navratil2009:ncsm} permits an exact factorization of
center-of-mass and intrinsic wave functions.\footnote{The $\Nmax$ truncation
is to many-body states with $0\leq\Nex\leq\Nmax$.  For NCSM
calculations with parity-conserving interactions,
the two parity subspaces, obtained for $\Nex$ even or odd, respectively, 
may furthermore be considered separately.}  This property also implies that the separation of CMF states may be carried
out separately within the space of states with each specific number of
oscillator quanta, which we denote by $W_\Nex$.  Furthermore, $\Hcm$
commutes with the $\grpsu{3}$ generators and may therefore be
diagonalized within a subspace with good $\grpsu{3}$ quantum
numbers~\cite{verhaar1960:shell-com}.  As an operator acting only on
spatial degrees of freedom, $\Hcm$ also commutes with the total spin
operators for protons and neutrons, as well as their combined spin operator.
Thus, collecting these properties, the separation of CMF states may be
carried out separately within subspaces of given $\Nex$,
$(\lambda\mu)$, $S_{\proton}$, $S_{\neutron}$, and $S$, which we
denote by $W_{\Nex}[(\lambda\mu)S_{\proton}S_{\neutron}S]$.

In considering how to extract CMF states, we note that these states
are defined by their relation to the center-of-mass harmonic
oscillator raising and lowering operators.  The
center-of-mass raising operator $\inline{\Bdcm}$, which is an $L=1$ operator, furthermore constitutes an $\grpsu{3}$  $(10)$
tensor, with components $\inline{\Bdcmlm[L=1,M]}$.  This operator may be written in terms of
single-particle harmonic-oscillator raising operators
$b^{\dagger (10)}$
as~\cite{escher1998:sp6r-shell-su3coupling,dytrych2008:sp-ncsm}
\begin{equation}
\Bdcmlm=\frac{1}{\sqrt{A}}\sum_{i=1}^A
{b}^{\dagger (10)}(i).
\end{equation}
The corresponding lowering operator $\inline{\Btcmlm}$ similarly has the form
\begin{equation}\label{eqn:B_CM}
\Btcmlm = \frac{1}{\sqrt{A}} \sum_{i=1}^A
\tilde{b}^{(01)}(i).
\end{equation}
The center-of-mass harmonic oscillator Hamiltonian is
built from these operators as
\begin{equation}
\Hcm = \hbar\omega\Big(\Bdcmlm\cdot\Btcmlm + \frac{3}{2}\Big),
\end{equation}
where the dot indicates a spherical tensor scalar product.
The CMF states are then defined by the property that they have no
center-of-mass excitations, \textit{i.e.}, they have zero eigenvalue
for the center-of-mass number operator
$\inline{\Ncm=\Bdcmlm\cdot \Btcmlm}$.
Equivalently, however, they are identified by the property that they
are annihilated by the center-of-mass
lowering operator $\inline{\Btcmlm}$.

Either of these criteria allow the problem of identifying CMF states
to be formulated as a null-space problem.  We seek the subspace
$W^{\mathrm{CMF}}_{\Nex}[(\lambda\mu)S_{\proton}S_{\neutron}S]$ of
$W_{\Nex}[(\lambda\mu)S_{\proton}S_{\neutron}S]$ consisting of states
$\tket{\Psi^\mathrm{CMF}}$ such that $\Ncm\tket{\Psi^\mathrm{CMF}}=0$ or, equivalently,
$\inline{\Btcmlm[L=1,M]\tket{\Psi^\mathrm{CMF}}=0}$.  In practice, this means
first representing the operator as a matrix with respect to the
$\grpsu{3}$-NCSM basis and then solving for a complete set of null vectors of this
matrix.  The result yields new basis states, for
$W^{\mathrm{CMF}}_{\Nex}[(\lambda\mu)S_{\proton}S_{\neutron}S]$, as
linear combinations of the original basis states, for
$W_{\Nex}[(\lambda\mu)S_{\proton}S_{\neutron}S]$.

Although we could in principle search for the null space of either
$\Ncm$ or $\inline{\Btcmlm}$, there is a practical advantage to
working with $\inline{\Btcmlm}$.  While $\Ncm$ is a two-body operator,
$\inline{\Btcmlm}$ is simply a one-body operator. Consequently,
evaluation of matrix elements is computationally less involved.  Note
that $\Ncm$ acts within the space $W_\Nex$, \textit{i.e.}, conserving
the number of oscillator excitations, but $\inline{\Btcmlm}$ connects
the space $W_\Nex$ to the next lower space $W_{\Nex -1}$, which is
significantly smaller in dimension than $W_\Nex$. One might therefore
expect the null space problem for $\inline{\Btcmlm}$ to be of lower
dimensionality than that for $\Ncm$, namely, involving a matrix of the
same column dimension ($\sim\dim W_{\Nex }$) but lower row dimension
($\sim\dim W_{\Nex -1}$). However, this simple relation of dimensions
is modified once $\grpsu{3}$ selection rules are considered (see
below), and in practice the difference in dimensionality of the two
problems is largely eliminated.

The problem of identifying CMF states is simplified by the realization that it may be formulated
entirely at the level of reduced states.  Recall that $\Hcm$ commutes with the
$\grpsu{3}$ generators, which connect states within an irrep.  Consequently, an
$\grpsutimes$-reduced state is CMF if and only if its substates,
labeled by $\kappa L J$ (and $M$), are all CMF.    We need
thus only find a basis of $\grpsutimes$-\textit{reduced} states, for
$W^{\mathrm{CMF}}_{\Nex}[(\lambda\mu)S_{\proton}S_{\neutron}S]$,
independent of $\kappa L JM$.  We also need only consider the
$\grpsutimes$-\textit{reduced} matrix elements of the operator
$\inline{\Btcmlm}$, rather than the matrix
elements among individual $\kappa L JM$ substates, which are much
greater in number.

To explicitly relate the null space problem at the level of individual states to that for
$\grpsutimes$-reduced matrix elements, let us return for a moment to
$\kappa LJM$ states and observe that any CMF state
$\tket{\Psi^{\mathrm{CMF}}_{\Nex}[(\lambda\mu)S_{\proton}S_{\neutron}S;\kappa LJM]}$ within the $W_{\Nex}[(\lambda\mu)S_{\proton}S_{\neutron}S]$
subspace must satisfy
\begin{equation}\label{eqn:B_CM_matrix_element}
\tme{
\Psi_{\Nex-1}[(\lambda^{\prime}\mu^{\prime})S^{\prime}_{\proton}S^{\prime}_{\neutron}S^{\prime};\kappa^{\prime}
L^{\prime}J^{\prime}M^{\prime}]
}{
\Btcmlm[L_0=1,M_0]
}{
\Psi^{\mathrm{CMF}}_{\Nex}[(\lambda\mu)S_{\proton}S_{\neutron}S;\kappa
LJM] 
} = 0,
\end{equation}
for every state 
$\tket{\Psi_{\Nex-1}[(\lambda^{\prime}\mu^{\prime})S^{\prime}_{\proton}S^{\prime}_{\neutron}S^{\prime};\kappa^{\prime}
L^{\prime}J^{\prime}M^{\prime}]}\in
W_{\Nex-1}$.  Since we are working with states of good angular
momentum, we can immediately rewrite the condition in terms of a
reduced matrix element as 
\begin{equation}\label{eqn:B_CM_matrix_element_SU2}
\trme{
\Psi_{\Nex-1}[(\lambda^{\prime}\mu^{\prime})S^{\prime}_{\proton}S^{\prime}_{\neutron}S^{\prime};\kappa^{\prime}L^{\prime}J^{\prime}]
}{
\Btcmlm[L_0=1]
}{
\Psi^{\mathrm{CMF}}_{\Nex}[(\lambda\mu)S_{\proton}S_{\neutron}S;\kappa
LJ] 
} = 0.
\end{equation}
Note that, since
$\inline{\Btcmlm}$ acts only on spatial degrees of freedom, we
actually need
only consider the case
$(S^{\prime}_{\proton}S^{\prime}_{\neutron}S^{\prime})=(
S_{\proton}S_{\neutron}S)$.  The ordinary $\grpsu{2}$-reduced matrix element
in~(\ref{eqn:B_CM_matrix_element_SU2}) is related to the $\grpsutimes$-reduced matrix element of $\inline{\Btcmlm}$ by the $\grpsu{3}$
Wigner-Eckart theorem (and $LS$-coupling relations), as
\begin{multline}\label{eqn:B_CM_wigner_eckart}
\trme{
\Psi_{\Nex-1}[(\lambda^{\prime}\mu^{\prime})S_{\proton}S_{\neutron}S;\kappa^{\prime}L^{\prime}J^{\prime}]
}{
\Btcmlm[L_0=1]
}{
\Psi^{\mathrm{CMF}}_{\Nex}[(\lambda\mu) S_{\proton}S_{\neutron}S;\kappa
LJ] 
}  
\\
=
(-)^{J+L'+1+S}\hat{L}' \hat{J}'\sixj{L}{J}{S}{J'}{L'}{1}
 \rcc{(\lambda\mu)}{\kappa
L}{(01)}{1}{(\lambda^{\prime}\mu^{\prime})}{\kappa^{\prime}L^{\prime}}
\\
\times
\trme{
\Psi_{\Nex-1}[(\lambda^{\prime}\mu^{\prime})S_{\proton}S_{\neutron}S]
}{
\Btcmlm
}{
\Psi^{\mathrm{CMF}}_{\Nex}[(\lambda\mu)S_{\proton}S_{\neutron}S] 
},
\end{multline}
where $\hat{J}\equiv (2J+1)^{1/2}$, and the quantity in parentheses is an $\grpsu{3}$ Clebsch-Gordan
coefficient~\cite{akiyama1973:su3-cg}.
The condition that the complete set of $\grpsu{2}$-reduced matrix elements appearing on
the left-hand side of~(\ref{eqn:B_CM_wigner_eckart}) vanish is
equivalent to the condition that the single $\grpsutimes$-reduced matrix element on
the right-hand side vanish.

Thus, for the CMF states, we seek $\grpsutimes$-reduced states $\trket{\Psi^{\mathrm{CMF}}_{\Nex}[(\lambda\mu);
S_{\proton}S_{\neutron}S]}$ such that
\begin{equation}\label{eqn:B_CM_reduced_matrix_element}
\trme{\Psi_{\Nex-1}[(\lambda^{\prime}\mu^{\prime})S_{\proton}S_{\neutron}S]}{\Btcmlm}{\Psi^{\mathrm{CMF}}_{\Nex}[(\lambda\mu)S_{\proton}S_{\neutron}S] }=0,
\end{equation}
for all possible reduced states
$\trket{\Psi_{\Nex-1}[(\lambda^{\prime}\mu^{\prime})S_{\proton}S_{\neutron}S]}
\in W_{\Nex-1}[(\lambda^{\prime}\mu^{\prime})S_{\proton}S_{\neutron}S]$.
The 
subspaces
$W_{\Nex-1}[(\lambda^{\prime}\mu^{\prime})S_{\proton}S_{\neutron}S]$ which
may be linked with 
$W_{\Nex}[(\lambda\mu)S_{\proton}S_{\neutron}S]$ through
$\inline{\Btcmlm}$ are restricted by the
$\grpsu{3}$ tensor character of $\inline{\Btcmlm}$.  Specifically, $(\lambda^{\prime}\mu^{\prime})$ must be contained in the
product $(\lambda\mu)\times (01)$ which, from the general rules of
$\grpsu{3}$ coupling~\cite{oreilly1982:su3-coupling}, may be
seen to consist of
\begin{equation}\label{eqn:selection_rules}
(\lambda\mu)\times (01)= \begin{cases}
(01) &\lambda=\mu=0  \\
(0~\mu+1)\oplus (1~\mu-1) & \lambda=0,\mu \ge 1 \\ 
(\lambda 1)\oplus (\lambda-1~0) & \lambda\ge 1, \mu=0\\
(\lambda~\mu+1 )\oplus(\lambda+1~\mu-1)\oplus (\lambda-1~\mu) &
\lambda\ge1, \mu\ge1
.
			 \end{cases}
\end{equation}

The problem of finding reduced states which
satisfy~(\ref{eqn:B_CM_reduced_matrix_element}) can be converted into
searching for the null space of a matrix, the entries of which are the
$\grpsutimes$-reduced matrix elements of $\inline{\Btcmlm}$ between the
$\grpsutimes$-reduced basis states for the
$W_{\Nex}[(\lambda\mu)S_{\proton}S_{\neutron}S]$ subspace and the
$\grpsutimes$-reduced basis states for each of the possible subspaces
$W_{\Nex-1}[(\lambda^{\prime}\mu^{\prime})S_{\proton}S_{\neutron}S]$.
The resulting matrix has the form illustrated in
Fig.~\ref{fig-matrix}, where the horizontal dashed lines delimit
submatrices corresponding to the different final spaces with
$(\lambda'\mu')=(\lambda_1'\mu_1')$, $(\lambda_2'\mu_2')$, $\ldots$,
as allowed by the selection rule~(\ref{eqn:selection_rules}). (There will
be at most three such submatrices.) The entries of the null
vectors then give the expansion coefficients for the basis states for
the CMF space
$W^{\mathrm{CMF}}_{\Nex}[(\lambda\mu)S_{\proton}S_{\neutron}S]$ in
terms of the original basis states of
$W_{\Nex}[(\lambda\mu)S_{\proton}S_{\neutron}S]$.
%%%%%%%%%%%%%%%%%%%%%%%%%%%%%%%%%%%%%%%%%%%%%%%%%%%%%%%%%%%%%%%%
\begin{figure}
\begin{center}
%% \Mentry{bralmindex}{braindex}{ketindex}
\newcommand{\Mentry}[3]{\trme{\Psi^{(\lambda_#1'\mu_#1')}_{\Nex-1,#2}}{\Btcmlm}{\Psi^{(\lambda\mu)}_{\Nex,#3}}}
\begin{math}
\pmatset{3}{4em}  % intercolumn separation
\begin{pmat}[{..}]
\Mentry{1}{1}{1} & \Mentry{1}{1}{2} & \cdots \cr
\Mentry{1}{2}{1} & \Mentry{1}{2}{2} & \cdots \cr
\vdots & \vdots  &  \cr\-
\Mentry{2}{1}{1} & \Mentry{2}{1}{2} & \cdots \cr
\Mentry{2}{2}{1} & \Mentry{2}{2}{2} & \cdots \cr
\vdots & \vdots  &  \cr\-
\vdots & \vdots 
\cr
\end{pmat}
\end{math}
\end{center}
~\\[-24pt]
\caption{Form of the matrix representation of
$\Btcmlm$, as reduced matrix elements  
between the $\grpsu{3}$-NCSM basis states for the
$W_{\Nex}[(\lambda\mu)S_{\proton}S_{\neutron}S]$ subspace (identified
with the columns) and the $\grpsu{3}$-NCSM basis states for each of the possible subspaces
$W_{\Nex-1}[(\lambda^{\prime}\mu^{\prime})S_{\proton}S_{\neutron}S]$
(identified with the rows).
}
\label{fig-matrix}
\end{figure}
%%%%%%%%%%%%%%%%%%%%%%%%%%%%%%%%%%%%%%%%%%%%%%%%%%%%%%%%%%%%%%%%

The $\grpsu{3}$-reduced matrix
elements of $\inline{\Btcmlm}$ entering into
the matrix of Fig.~\ref{fig-matrix}
can be calculated numerically using methods
from Ref.~\cite{bahri1994:su3-rme}. In particular, the $\grpsutimes$ matrix
elements of a one-body operator such as $\inline{\Btcmlm}$ (or
an $n$-body operator in general) can readily be computed, once the
operator is expressed in second-quantized form in terms of
$\grpsutimes$-coupled products of creation operators
$\inline{a^{\dagger}_{(\Nsp 0)1/2}}$ and annihilation operators $\inline{\tilde{a}_{(0n)1/2}}$~\cite{escher1998:sp6r-shell-su3coupling}.
For the center-of-mass annihilation
operator $\inline{\Btcmlm}$, the second-quantized form is
obtained as
\begin{equation}\label{eqn:b_second_quantized}
\Btcmlm 
=
\frac{1}{\sqrt{A}}
 \sum_{\Nsp} 
\sqrt{\frac{(\Nsp+1)(\Nsp+2)}{3}}\trme{\Nsp}{\tilde{b}^{(01)}}{\Nsp+1} \left[
a^{\dagger}_{(\Nsp0)1/2} \times \tilde{a}_{(0,\Nsp+1)1/2}
\right]^{(01)0},
\end{equation}
where $\trme{\Nsp}{\tilde{b}^{(01)}}{\Nsp+1}=\sqrt{\Nsp+3}$.

%%%%%%%%%%%%%%%%%%%%%%%%%%%%%%%%%%%%%%%%%%%%%%%%%%%%%%%%%%%%%%%%

\section{Dimensions of CMF spaces}
\label{sec-calc}

Let us now examine the dimensions of the CMF subspaces obtained by the
methods of Section~\ref{sec-cmf}.  Our primary interest is in the
distribution of CMF and spurious states with respect to the
$\grpsutimes$ quantum numbers.  The Sp-NCSM approach requires
identification of these CMF $\grpsutimes$ irreps, at low $\Nex$, from which
$\grpsptr$ irreps extending to high $\Nex$ are created by repeated
action of the symplectic raising operator (Section~\ref{sec-intro}).
%%%%%%%%%%%%%%%%%%%%%%%%%%%%%%%%%%%%%%%%%%%%%%%%%%%%%%%%%%%%%%%%
\begin{figure}
\begin{center}
\includegraphics[width=\hsize]{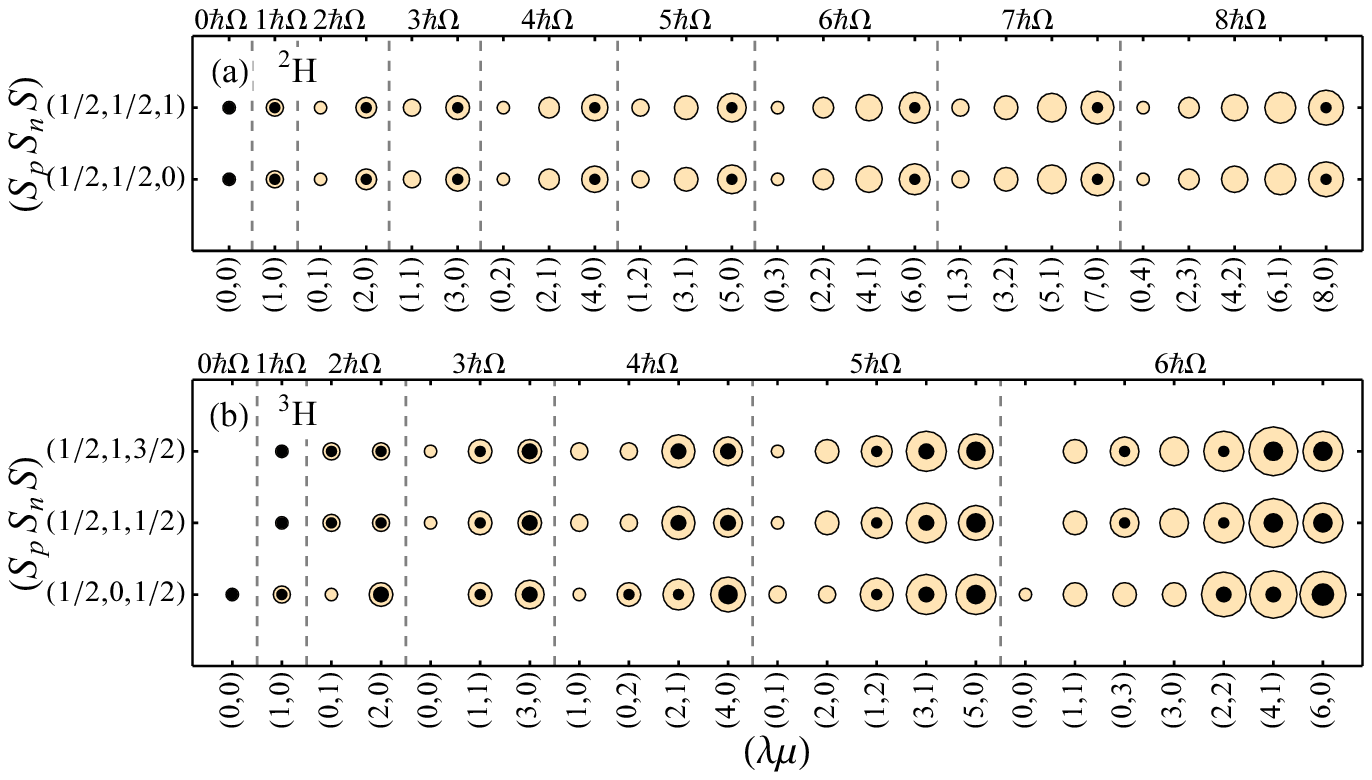}
\end{center}
~\\[-36pt]
\caption{Dimensions of the $\grpsutimes$-reduced subspaces of the
$\grpsu{3}$-NCSM model space for (a)~the
deuteron ($\isotope[2]{H}$) and (b)~the
triton ($\isotope[3]{H}$).  
The area of the light (outer) circle indicates the full dimension,
\textit{i.e.}, of
$W_{\Nex}[(\lambda\mu)S_{\proton}S_{\neutron}S]$, and the area of the
dark (inner) circle indicates the dimension of its CMF subspace,
\textit{i.e.}, of
$W^{\mathrm{CMF}}_{\Nex}[(\lambda\mu)S_{\proton}S_{\neutron}S]$.
Subspaces are arranged by $\Nex$ (labeled by $\Nex\hbar\Omega$ at
top), $\grpsu{3}$ quantum numbers (at bottom), and spin quantum
numbers $(S_{\proton}S_{\neutron}S)=(\tfrac12\tfrac12S)$ (at left).  }
\label{fig-cmf-dim-2h-3h}
\end{figure}
%%%%%%%%%%%%%%%%%%%%%%%%%%%%%%%%%%%%%%%%%%%%%%%%%%%%%%%%%%%%%%%%

To begin with, the simplest illustration we might consider is the
model space for the deuteron ($\isotope[2]{H}$), which is shown along
with that for the triton ($\isotope[3]{H}$) in
Fig.~\ref{fig-cmf-dim-2h-3h}.  The full $\grpsu{3}$-NCSM space for
the deuteron may be broken into subspaces $W_{\Nex}[(\lambda\mu)
\frac{1}{2} \frac{1}{2} S]$, with $S=0$ and $1$.  The dimensions of
these subspaces~--- by which we specifically mean the
number of $\grpsutimes$-reduced basis states, not the total number of
$\kappa LJ$ basis states, which would be much higher~--- are 
indicated by the areas of the outer, light circles in
Fig.~\ref{fig-cmf-dim-2h-3h}.    
Then, for each
value of $\Nex$, it is found that solution of the null space problem for the center-of-mass annihilation operator  yields two
CMF reduced states.  Both have
$(\lambda\mu)=(\Nex0)$, one with $S=0$ and one with $S=1$.  These CMF subspaces are 
indicated by the inner, dark circles in
Fig.~\ref{fig-cmf-dim-2h-3h}. 

The deuteron provides a particularly illuminating example, since the
quantum numbers of the CMF spaces obtained in
Fig.~\ref{fig-cmf-dim-2h-3h} may  be understood through simple
arguments.  Let $\Ncm$ and $\Nrel$ denote the number of oscillator
excitations of the center-of-mass and relative degrees of freedom,
respectively, so $\Nex = \Ncm + \Nrel$.  The CMF condition imposes $\Ncm = 0$ and thus $\Nrel = \Nex$.
For the two-particle system, the
transformation between single-particle and relative coordinates is
straightforward. There is only a single relative
coordinate vector, and the harmonic oscillator in this 
coordinate is equivalent to the
harmonic oscillator for a
single particle in three dimensions.  For a given $\Nex$, it will therefore carry
$\grpsu{3}$ quantum numbers $(\Nex0)$.  The center-of-mass
oscillator carries $(00)$ for a CMF state. Therefore the CMF state as
a whole will have
$(\lambda\mu)=(\Nex0)\times(00)=(\Nex0)$, as well.  
Since the deuteron consists of distinguishable particles, the coupling of spins is independent of the spatial
degrees of freedom, and both $S=0$ and $1$ are obtained.

\begin{figure}
\begin{center}
\includegraphics[width=\hsize]{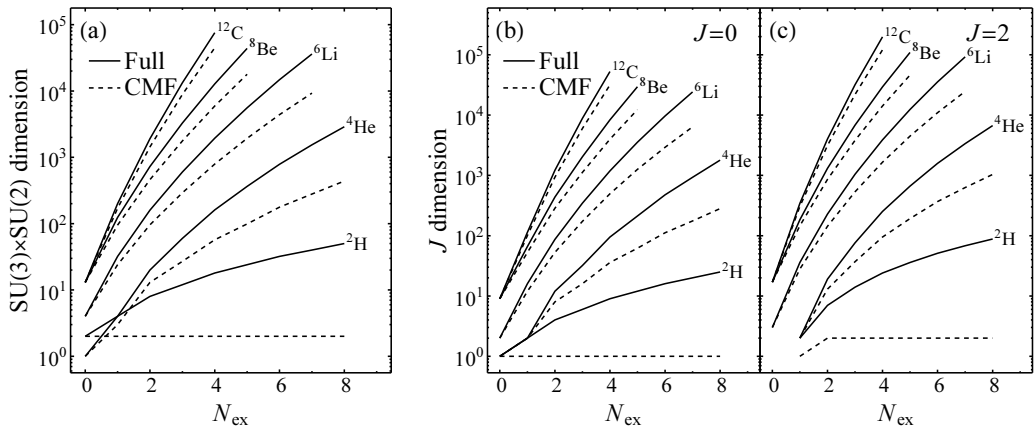}
\end{center}
~\\[-36pt]
\caption{Dimension of the 
$\grpsu{3}$-NCSM model space, decomposed according to $\Nex$: (a)~by
number of $\grpsutimes$-reduced basis states,
(b)~after branching to $J=0$ states, and (c)~after branching to $J=2$
states.  Both full (solid curves) and CMF (dashed curves) dimensions
are shown, for $\isotope[2]{H}$, $\isotope[4]{He}$, $\isotope[6]{Li}$,
$\isotope[8]{Be}$, and $\isotope[12]{C}$.  
}
\label{fig-dim-cmf-combo}
\end{figure}

In general, as we move beyond the two-body problem, the fraction of
the model space dimension which corresponds to CMF states, versus that
which corresponds to spurious states, depends both on the number of
nucleons and on the number $\Nex$ of oscillator quanta.  The spurious
contribution at $\Nex=0$ is identically vanishing, as
shown by Elliott and Skyrme~\cite{elliott1955:com-shell}.  For a given nucleus, the spurious contribution becomes an
increasing fraction of the total space with increasing $\Nex$, while,
for a given level of
excitation $\Nex$, the spurious contribution becomes a less significant
fraction of the total space  with
increasing atomic mass $A$.  
These trends are already in evidence in comparing deuterium
with tritium (Fig.~\ref{fig-cmf-dim-2h-3h}) but may be seen more
systematically in Fig.~\ref{fig-dim-cmf-combo}(a), where
the dimensions of the full spaces (solid curves) and CMF spaces
(dashed curves) are shown, as functions of $\Nex$, for several nuclei
with $2\leq A \leq 12$.  
%%%%%%%%%%%%%%%%%%%%%%%%%%%%%%%%%%%%%%%%%%%%%%%%%%%%%%%%%%%%%%%%
%%%%%%%%%%%%%%%%%%%%%%%%%%%%%%%%%%%%%%%%%%%%%%%%%%%%%%%%%%%%%%%%

Qualitatively, the pattern in the evolution of the CMF fraction may be understood from simple counting
arguments, by considering the
possible ways of distributing $\Nex$ oscillator quanta over the $3A$
oscillator degrees of freedom of the nuclear system.  These may be decomposed into three
center-of-mass oscillator degrees of freedom and $3A-3$ relative oscillator
degrees of freedom.  If many oscillator quanta are to be distributed
over few degrees of freedom ($\Nex\gg A$), most of the distributions will allocate
at least one oscillator quantum to the center-of-mass degrees of
freedom, leading to a high proportion of spurious
states, \textit{i.e.}, a smaller CMF fraction.  Conversely,
if few
oscillator quanta are to be distributed over many degrees of freedom
($\Nex\ll A$), most distributions will ``miss'' these three center-of-mass degrees of
freedom, \textit{i.e.}, \textit{not} allocate any
oscillator quanta to them, leading to a low proportion of spurious
states, \textit{i.e.}, a larger CMF fraction.\footnote{Quantitatively,
such counting
arguments give an
estimated CMF fraction
$(\dim W_\Nex^\mathrm{CMF})/(\dim W_\Nex)\linebreak[0]\approx\inline{\falling{(3A-1)}{3}/\linebreak[0]\falling{(\Nex+3A-1)}{3}}$,
where $\falling{m}{n}\equiv m(m-1)\cdots(m-n+1)$.  This expression
gives the exact result for the deuteron, when applied to the total
space, \textit{i.e.}, counting all $M$ substates, but is of only approximate validity
for other nuclei, due to neglect of antisymmetrization.  For large
$\Nex$, we obtain a CMF fraction falling as
$\NexBARE^{-3}$.}

%%%%%%%%%%%%%%%%%%%%%%%%%%%%%%%%%%%%%%%%%%%%%%%%%%%%%%%%%%%%%%%%
\begin{figure}[p]
\begin{center}
\includegraphics[width=\hsize]{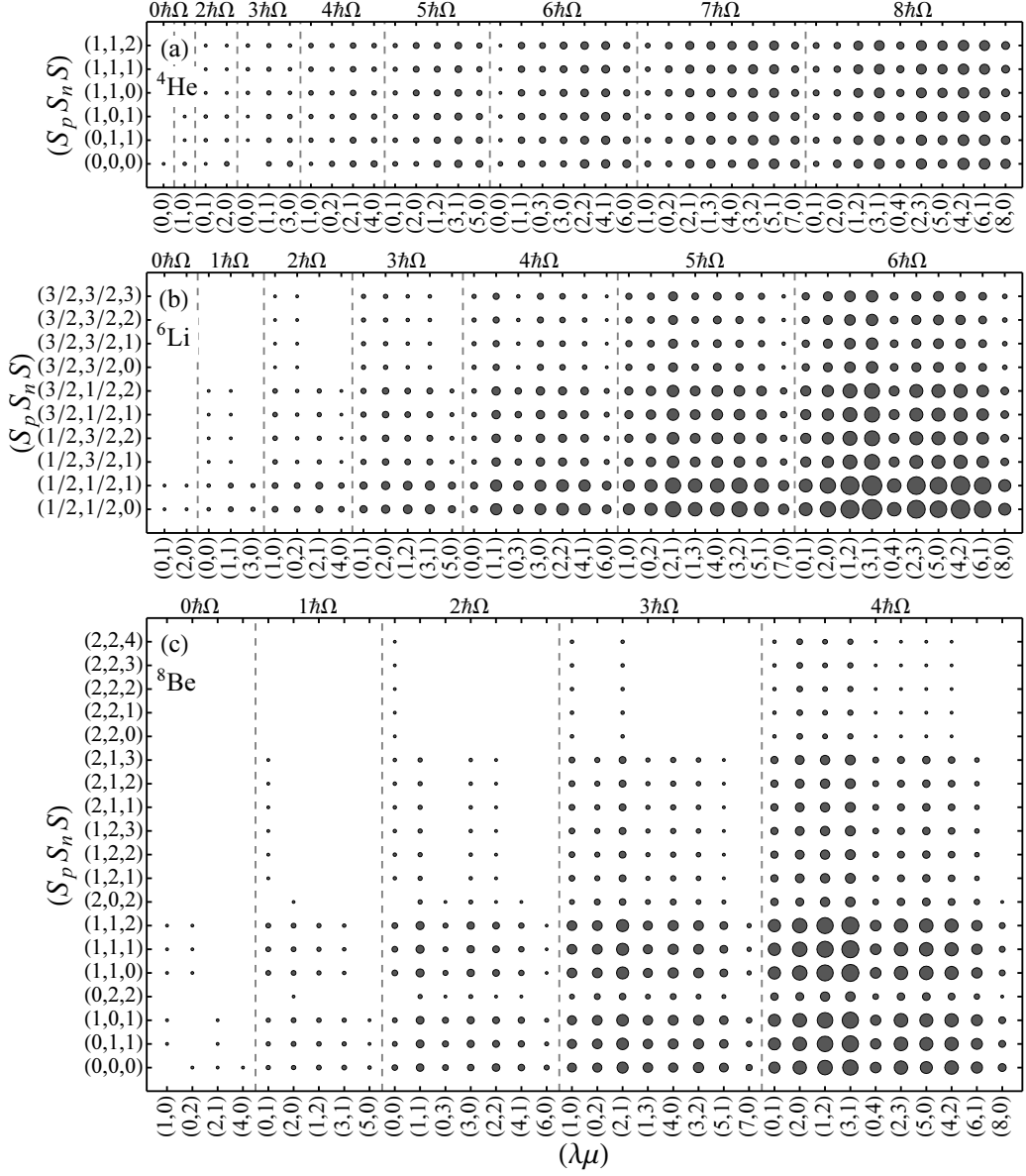}
\end{center}
~\\[-36pt]
\caption{Dimensions of the $\grpsutimes$-reduced subspaces for
(a)~$\isotope[4]{He}$, (b)~$\isotope[6]{Li}$, and
(c)~$\isotope[8]{Be}$.  The size of each circle represents the
dimension, scaled such that a doubling in radius
represents a tenfold increase in dimension.  Subspaces are arranged by
$\Nex(\lambda\mu)S_pS_nS$ as indicated in the caption to
Fig.~\ref{fig-cmf-dim-2h-3h}.  }
\label{fig-dim-combo}
\end{figure}
%%%%%%%%%%%%%%%%%%%%%%%%%%%%%%%%%%%%%%%%%%%%%%%%%%%%%%%%%%%%%%%%

%%%%%%%%%%%%%%%%%%%%%%%%%%%%%%%%%%%%%%%%%%%%%%%%%%%%%%%%%%%%%%%%
\begin{figure}[p]
\begin{center}
\includegraphics[width=\hsize]{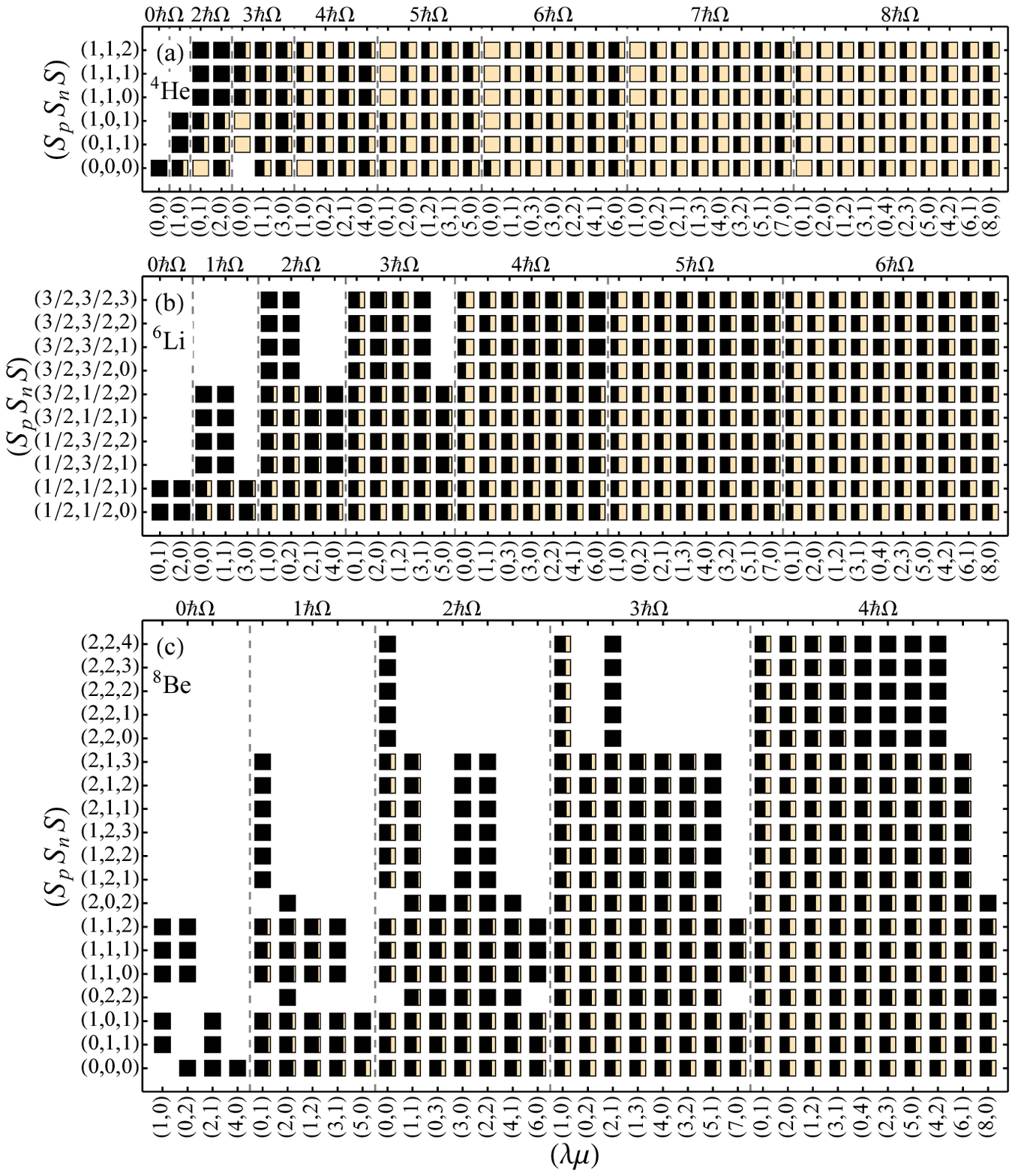}
\end{center}
~\\[-36pt]
\caption{Ratio of CMF dimension to total dimension, 
for the $\grpsutimes$-reduced subspaces for
(a)~$\isotope[4]{He}$, (b)~$\isotope[6]{Li}$, and (c)~$\isotope[8]{Be}$.  
The area of the darkened portion of each square, relative to the total area,
indicates the dimension of the CMF subspace, \textit{i.e.}, of
$W^{\mathrm{CMF}}_{\Nex}[(\lambda\mu)S_{\proton}S_{\neutron}S]$
relative to that of the full subspace, \textit{i.e.}, of
$W_{\Nex}[(\lambda\mu)S_{\proton}S_{\neutron}S]$.
Subspaces are arranged by $\Nex(\lambda\mu)S_pS_nS$ as indicated in the caption to Fig.~\ref{fig-cmf-dim-2h-3h}.
 }
\label{fig-cmf-ratio-combo}
\end{figure}
%%%%%%%%%%%%%%%%%%%%%%%%%%%%%%%%%%%%%%%%%%%%%%%%%%%%%%%%%%%%%%%%

We focus now on the detailed distribution of CMF and spurious states
over the $\grpsutimes$ subspaces of $W_\Nex$, that is, the
$\Nex(\lambda\mu)S_pS_nS$ subspaces of Section~\ref{sec-cmf}.  First,
it should be noted that the dimensions of these subspaces vary widely.
The dimensions of the $\grpsutimes$ subspaces for
$\isotope[4][2]{He}_2$, $\isotope[6][3]{Li}_3$, and
$\isotope[8][4]{Be}_4$, with $\Nex\leq8$, $6$, and $4$, respectively,
are shown in Fig.~\ref{fig-dim-combo}~--- these dimensions vary from
one to $640$. To accommodate this range, the circles in
Fig.~\ref{fig-dim-combo} have been sized according to a power law scale, such
that a doubling in radius represents a tenfold increase in dimension.
However, any such nonlinear rescaling precludes meaningful visual comparison
of the dimensions of the full and CMF subspaces, as was possible in
Fig.~\ref{fig-cmf-dim-2h-3h}.  Consequently, we examine the CMF
fractions, \textit{i.e.}, the ratio $(\dim
W^{\mathrm{CMF}}_{\Nex}[(\lambda\mu)S_{\proton}S_{\neutron}S])/(\dim
W_{\Nex}[(\lambda\mu)S_{\proton}S_{\neutron}S])$, separately in
Fig.~\ref{fig-cmf-ratio-combo}, for these same nuclei.

As already noted, the CMF fractions decrease with increasing $\Nex$
and approach unity, for a given $\Nex$, with increasing mass.
However, within each $\Nex$ space, demarcated by 
vertical dashed lines in Fig.~\ref{fig-cmf-ratio-combo}, a further dominant trend may be
noticed in the variation of CMF fractions among the $\grpsutimes$ subspaces.  It is observed
that the spurious states are preferentially found in the subspaces
with $\grpsu{3}$ quantum numbers corresponding to the lowest values of
the $\grpsu{3}$ second-order Casimir
invariant $C_2$,  while the CMF states are
preferentially found in the subspaces corresponding to the highest
eigenvalues of $C_2$.
In Fig.~\ref{fig-cmf-ratio-combo} (as well as in
Figs.~\ref{fig-cmf-dim-2h-3h} and~\ref{fig-dim-combo}), within each $W_\Nex$, the
$\grpsu{3}$ irrep labels $(\lambda\mu)$ are ordered from left to right
by increasing eigenvalue of $C_2$ (given by
$\tbracket{C_2}=\lambda^2+\lambda\mu+\mu^2+3\lambda+3\mu$), where
labels which are degenerate with respect
to $C_2$ are then ordered by increasing $\lambda$.  The lowest CMF
fractions are (predominantly) found at left, and the highest
(predominantly) at right.  While some evidence may be seen for
patterns in the distribution with respect to the spin labels, these
are not as clear or consistent.

So far, we have
considered dimensions at the level of $\grpsutimes$-reduced basis
states.
We may also
deduce from these the full and CMF dimensions in terms of $J$-coupled states. 
Although much of  the computational process in an $\grpsu{3}$-NCSM
calculation can be carried out in terms of 
$\grpsutimes$-reduced states, and it is these reduced states which are
relevant to the definition of the Sp-NCSM model space, the Hamiltonian
matrix in an $\grpsu{3}$-NCSM calculation must ultimately be realized
in terms of basis states
of definite angular
momentum $J$.  As outlined in Section~\ref{sec-su3}, the $J$ states are obtained by first branching
each $\grpsutimes$-reduced state~(\ref{eqn:SU3xSU2_basis_states}) to
states of definite orbital angular momentum $L$.  The $L$ values
contained within an $\grpsu{3}$ irrep
$(\lambda\mu)$ are given by the $\grpsu{3}\rightarrow\grpso{3}$ branching
rule~\cite{elliott1958:su3-part1}.\footnote{For each value $K=
\min(\lambda,\mu) \bmod
2,\dots,\min(\lambda,\mu)-2,\min(\lambda,\mu)$, states are obtained
with $L=K, K+1, \ldots,K+\max(\lambda,\mu)$, with the exception that
$L=\max(\lambda,\mu) \bmod 2,
\ldots,\max(\lambda,\mu)-2,\max(\lambda,\mu)$ if $K=0$.}  Then $L$ is coupled
with $S$ to yield states of total angular momentum $J$ according to
the usual coupling rules for angular momentum addition.  The
dimensions of the resulting $J$-spaces~--- both the full space and its
CMF portion~--- are shown for several light nuclei in
Fig.~\ref{fig-dim-cmf-combo}(b,c), for $J=0$ and $J=2$,
respectively.  Although the dimensions of the $J$-spaces are
calculated here via the $\grpsutimes$ coupling scheme, the results
obtained are generally applicable to any $J$-coupled scheme for the
NCSM, in an $\Nmax$ truncation.

%%\clearpage
\section{Conclusions}

The separation or elimination of spurious center-of-mass excitations is essential to
the problem of determining the intrinsic structure of nuclei.  If an
$\grpsu{3}$-coupled harmonic oscillator basis is used for the
many-body problem, as in the $\grpsu{3}$-NCSM, the separation may be
carried out at the level of $\grpsu{3}$ irreps, in particular, within
subspaces of fixed number of oscillator quanta and
$(\lambda\mu)S_{\proton}S_{\neutron}S$ labels.  We have formulated the
problem of finding the CMF subspace as a matrix null-space problem,
based on the $\grpsutimes$-reduced matrix elements of the
center-of-mass annihilation operator, which is solved
independently for each $\grpsutimes$ subspace of the full model
space. 
It is therefore possible to
remove spurious contributions from the $\grpsu{3}$-NCSM model space
\textit{prior} to diagonalization of the Hamiltonian, rather than
through a Lawson term.  In the context of the $\grpsu{3}$-NCSM
\textit{per se}, this
raises the possibility of substantial reductions in dimensionality of
the problem,
principally in the high $\Nex$ subspaces of lighter nuclei.  

However, a more essential application lies in providing the foundation
for ensuring exact separation of center-of-mass and intrinsic
dynamics, or removal of spurious contributions, in the Sp-NCSM.
The
purpose of the Sp-NCSM is to incorporate physically relevant portions
of the model space extending to much higher numbers of oscillator
quanta, beyond those which can be practically reached if one must retain
the complete
$\Nex(\lambda\mu)S_pS_nS$ subspaces of the $\grpsu{3}$-NCSM.
However, factorization of center-of-mass and intrinsic wave
functions is still guaranteed if the $\grpsptr$ basis is constructed
starting from $\grpsu{3}$-NCSM extremal states which are free of
spurious excitation.
The present approach
may also serve as the starting point for eliminating spurious
admixtures in other extensions to the $\grpsu{3}$-NCSM, such as
an adaptation of the importance truncation
scheme~\cite{roth2009:it-ncsm-16o} to the $\grpsu{3}$-NCSM.

%%\ack 
\section*{Acknowledgements}
Discussions with C.~Bahri are gratefully acknowledged. This
work was supported by the Research Corporation for Science
Advancement under a Cottrell Scholar Award, by the US Department
of Energy under Grants No.~DE-FG02-95ER-40934  and DE-SC0005248, and by the
US National Science Foundation under Grant No.~OCI-0904874.
Computational resources were provided by the University of Notre
Dame Center for Research Computing.

%% \vfil % omit for NPA online submission

%%\section*{References} % omit for NPA online submission

% Bibliography created with apsrevm.bst
\providecommand{\APSLONG}{}
\providecommand{\ELSEVIER}{}
\ELSEVIER

%bibliography{apsrevm_elsevier,master,mc,theory,expt,books,misc,su3cmf,proc}

\end{document}